# Fine-grained Mining of Illicit Drug Use Patterns Using Social Multimedia Data from Instagram

Yiheng Zhou, Numair Sani and Jiebo Luo
Department of Computer Science
University of Rochester
Rochester, New York 14627, USA
{yiheng.zhou, numair.sani, jiebo.luo}@rochester.edu

*Abstract*—According to NSDUH (National Survey on Drug Use and Health), 20 million Americans consumed drugs in the past few 30 days. Combating illicit drug use is of great interest to public health and law enforcement agencies. Despite of the importance, most of the existing studies on drug uses rely on surveys. Surveys on sensitive topics such as drug use may not be answered truthfully by the people taking them. Selecting a representative sample to survey is another major challenge. In this paper, we explore the possibility of using big multimedia data, *including both images and text*, from social media in order to discover drug use patterns *at fine granularity with respect to demographics*. Instagram posts are searched and collected by drug related terms by analyzing the hashtags supplied with each post. A large and dynamic dictionary of frequent drug related slangs is used to find these posts. User demographics are extracted using robust face image analysis algorithms. These posts are then mined to find common trends with regard to the time and location they are posted, and further in terms of age and gender of the drug users. Furthermore, by studying the accounts followed by the users of drug related posts, we extract common interests shared by drug users.

## I. INTRODUCTION

Traditionally, online surveys, hard-copy surveys and telephone surveys are used to conduct studies related to risky behavior. However, surveys are not only time-consuming but may also suffer from sampling noise and small scale [4]. Therefore, we need to find a new and effective method for studying risky behaviors. As of 2015, there are approximately 2.2 billion social media users [5]. With the growing availability of the Internet, this number is only going to grow further. Social media is steadily becoming a more mainstream part of our lives and contains large amounts of invaluable data that can be analyzed. The NSDUH, a national wide survey that aims to collect data on people's drug consumption habits, only had 67,800 respondents, which is insignificant compared to the number of social media users in the country. As social media becomes more popular, people start posting on every facet of life on them. This led us to choose social media as our choice of data source for this study. Drug abuse is a major problem for society, and the menace has been steadily growing. According to the National Institute of Drug Abuse, "substance abuse costs our society more than 484 billion dollars a year," which is about thrice of what we spend on cancer. In 2000, there were about 460,000 deaths caused by illicit drug abuse [6]. Therefore, drug abuse presents a real threat to society and we need more effective methods to combat such a threat. In this paper, we propose a novel approach to studying drug using patterns and networks by mining social media.

The main contributions of this study are as follows:
1. Employing Instagram as a reliable and large-scale data source to study patterns and trends related to major illegal drugs;
2. Applying item set mining algorithms to understand new linguistic trends related to drug culture;
3. Incorporating robust image analysis algorithms to extract drug user demographics;
4. Examining drug-user accounts to discover common interests shared by drug users;
5. Understanding the different networking behaviors of drug users (including drug dealers) and non-drug users.

## II. RELATED WORK

A number of studies on mining social media to discover certain patterns have been conducted by researchers, including predicting depression via social media [2], identifying disease outbreak by analyzing tweets [7], catching the criminals using Facebook [8], and analyzing alcohol-related promotions on Twitter [9]. In particular, Buntain and Golbeck [3] mined tweets from Twitter to discover drug use patterns. It is noteworthy to contrast our work with this particular work. For drug use time patterns, while their paper focuses on drug use time pattern based on days (100 days from April to July), our work can extract fine-grained time patterns in terms of day of week and hour of day. For location patterns, while their paper focuses on each state in United States, our work discovers the drug use distribution at the fine granularity of city level. More importantly, we employ the Face API of Project Oxford by Microsoft to study the age and gender patterns of drug users. Finally, we have developed a technique that could allow us to keep updating our database of drug-related hashtags because drug related lingo evolves.

## III. DATA SOURCE

Instagram is the choice of social media for this study because of its ease of use and high accuracy. It also has a massive, constantly expanding user base, which in turn supplies us with more data to analyze. In a recent announcement, Instagram announced that it has a user community of more than 400 million people [17] and to us, each one of these 400 million users is a potential drug user. Instagram is relatively anonymous, only requiring a nickname for an account and an email ID [4]. In contrast, Facebook requires more legitimate details. Moreover, teenagers often friend their parents on Facebook and thus are more selective in their posts [12]. Facebook has a strict real name policy, whereas Instagram has no such policy and allows users to register accounts under any alias.

More importantly, Instagram revolves around posting images about one's activities, and posts are usually accompanied by hashtags, single words that describe the activities/characteristics, which are what we use to fetch posts related to drug consumption. Based on that, we can easily fetch drug-related posts by searching for commonly used hashtags for drugs.

Moreover, Instagram has a highly active user base with the average number of posts of 2.69 posts per day and 49% of all the Instagram users use Instagram daily [14], plus about 80% of all the posts are talking about people themselves [31], which means people are willing to share their personal experience on Instagram including drug consumption [28]. Moreover, as the number of users of Instagram becomes bigger, people start doing business on Instagram by posting the products on the public pages, so do drug dealers [21]. Consequently, we can analyze drug users' behavior based on what people are posting under the influence.

## IV. METHODOLOGY

An important innovation in the methodology of this work is to leverage both image and text information gathered from a social media platform, such as Instagram, to facilitate fine-grained mining of illicit drug use patterns.

### A. Textual Analysis

We crawled the website for drug related posts. Instagram provides a RESTful API and we utilized this along with PHP's Guzzle library to make requests to the Instagram servers to identify and retrieve details of posts we deemed drug related. To identify the posts that were drug related, we examined the hashtags attached with the posts to see if the post contained any sensitive hashtags that we deemed drug associated. Specifically, we compiled a library of drug related hashtags, and these hashtags were identified using a dataset annotated by the New York State Attorney General's office. Initially, we obtained a collection of posts which have a drug-related hashtag "420". Each post contains hashtags, image, and comments on the post. 1,000 posts were annotated manually by experts as drug related. We manually selected the 100 most frequent drug related hashtags found in the 1000 posts. Equipped with these 100 hashtags, we performed a second round of search for posts containing these sensitive hashtags. Using the tag search API, we fetched 16,850,091 posts. Many of these posts are not drug-related but contain drug-related hashtags, so we filtered out the posts that have less than 2 drug-related hashtags. Next, we extracted all the unique usernames from the drug-related posts and obtained in total 2,362 potential drug users. With the list of usernames, Instagram user endpoint API was utilized to download all the posts from those 2362 users. Because we needed to infer demographics information from those users by applying facial identification API, we searched through all the posts for each user, counted the number of posts that have selfie-related hashtags ('selfie','weedselfie,' selfportrait' and 'selfy') and deleted the users that have fewer than two posts with selfie-related hashtags. Finally, 406 drug users were identified in our database. Sample posts are shown in Fig. 1.

Once we obtained the drug consumption activity posts from those 406 users, we extracted the username, user id, list of hashtags, time created and other attributes associated with the posts, and then used them for the subsequent analysis.

In this study, we only use hashtags in English hence it limits our posts to English speaking countries of the world that have a highly active Instagram population. Furthermore, most of these slangs are linked to the vocabulary of someone living on the North American or European Continent. Instagram stores the timestamp in GMT time, irrespective of the location of the posts, which presents itself as a source of error when we try to analyze peak times in a month drugs are consumed. Therefore, we searched through our database to find the posts with geolocations and measured the percentage of our users that are in United States (it is described in details in section V.B).

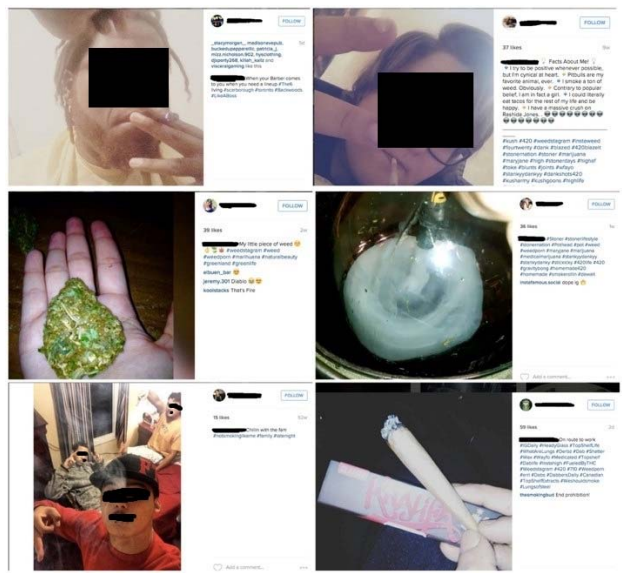

Figure 1. Sample drug related Instagram posts collected by Recent Media Endpoint (an API that Instagram provides).

## B. Image Analysis

Among the 406 sample drug users, we applied the facial recognition API to extract demographic data. For each post of the drug user in our database, we only took the posts with selfie related tags ('selfie','weedselfie,' selfportrait' and 'selfy') because we want to collect all the selfies of this user and calculate the face attributes (age and gender) by using Project Oxford. Additionally, even though some photos are tagged as 'selfie', they don't contain any faces. Therefore, by using the Detect API of Project Oxford, we can filter out the photos where no faces are detected.

In order to increase the accuracy of age detection, we averaged the ages obtained by mining selfies of those 406 users. Because, on average, each user has 32 selfies in our data set, we could have a standard deviation of less than 1-year-old (according to the standard deviation the declared by Project Oxford).

Finally, our data set contains some pictures (<1% of our data set) that have multiple faces. Because, usually, the photo taker's face is closest to the camera, we put this face into our database.

## C. Dynamic Linguistic Analysis

Since Instagram has been banning sensitive hashtags for drug use (such as '#dirtysprite'), drug-related hashtags are evolving. In order to keep our hashtag set up-to-date, we are using the Apriori algorithm [10] to mine frequently used hashtag sets. First, the initial hashtag set was used to fetch more drug related posts (according to the Instagram API policy, we can fetch more than 50,000 posts per hour and approximately 100 of them are considered as drug related) and since the information of post is stored in JSON format, we can easily obtain a list of hashtags used in this post by using Python Dictionary. Basically, we treated each post as a 'transaction' of Apriori and each hashtag as an 'item' in the 'transaction'. Then, we stored the data in CSV files with one 'transaction' each row and hashtags separated by comma. Finally, we applied our Apriori algorithm to those transactions and generated frequent item sets to update our hashtags dataset.

## D. Network Analysis

We then studied the network of the drug related community and compared it to that of normal Instagram users' community. For the drug related community, we utilized our 406 drug users as our initial nodes. For the normal Instagram users' community, we first used the tag search API to fetch posts that contain hashtag "#instapic" ("#instapic" is one of the most widely used hashtags by Instagram users. We need to obtain some random posts from Instagram, and extract the usernames from them to create a set of potential non-drug users). Next, we extracted the usernames from those posts and applied the user searching API to fetch all their posts. Finally, for each user, if none of his/her posts contains drug related hashtags in our database, we marked this account as non-drug related.

After having a decent number of users for both communities (Table I), we utilized the relationship endpoint of Instagram API to fetch the account information of all those users' followers and the accounts they follow. Next, we constructed two networks based on the follows/followed data for both drug related and non-drug related communities. Table II shows the number of nodes and edges for each network.

TABLE I. Number of Users from Sample Data

| Number of Drug Related Users | Number of Non-Drug Related Users |
|---|---|
| 406 | 1837 |

TABLE II. Information of drug-related and non-drug related networks

| Communities | Number of Nodes | Number of Edges |
|---|---|---|
| Drug Related | 308361 | 226374 |
| Non-Drug Related | 113321 | 185060 |

## E. Location Analysis

We also mined the patterns of drug uses based on the location. Instagram location endpoint can be used by specifying five parameters: minimum time, maximum time, distance (radius), longitude and latitude. First, we used a circle-making tool to make the searching areas cover the entire metropolitan area as shown in Fig. 2. Each circle represents the 5 parameters needed for location based search. For example, [1458220768, 1457615968, 5000, 33.740675, -118.260497] (the circle at the bottom) represents the request that will fetch all the posts within 5 kilometers centered at the Los Angeles International Airport from March 17th to March 24th. Even though some of the circles may have to overlap to cover all the areas, we have deleted all the duplications by image ID. We implemented a Python script to automatically fetch all the posts in those circles from year 2010 to 2015. Finally, we marked a post as drug-related if it contains more than 2 drug-related hashtags in our hashtag set. In total, we have fetched 1,232 drug-related posts in Los Angeles, CA, and 832 in Portland, OR. We chose Portland and Los Angeles because cannabis is legal in those two places, so it is more likely that people will show their locations when consuming drugs than any other places.

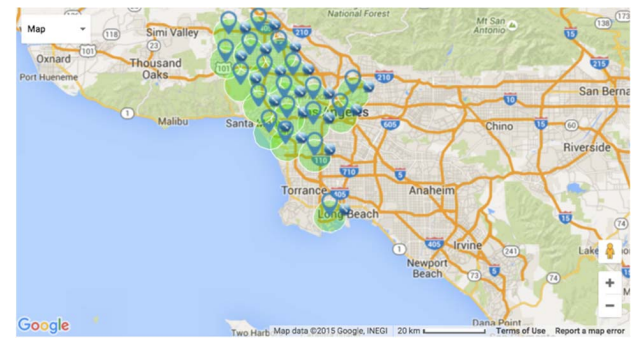

Figure 2. Circles used for location endpoint search in LA.

## V. RESULTS

### A. Popularity of Various Drugs

In this study, three classes of drugs are taken into consideration, namely weed (cannabis), cough syrup (purple drank), and prescription pills of various kinds like Vicodin that are frequent abused by people. Marijuana is the most popular choice of drug in United States and United Kingdom, as stated by the National Institute of Drug Abuse and the NHS, respectively. "It also comes in various forms called wax, a soft solid, oil, a gooey liquid and shatter, an amber like hard substance." Cough syrup is an over the counter drug that is abused by people. It is mixed with drinks in order to improve the flavor and its effects include relaxation and euphoria. Vicodin and other pills are regulated drugs that are sold at pharmacies with valid prescriptions and these are commonly abused by people. By searching through the drug users' posts and manually classifying cough syrup related, pills related and weed related hashtags, we are able to identify which drug activity this user was doing in the post by matching weed related, cough syrup related and pills related hashtags.

The result for drug popularity is in the Fig. 3. According to National Survey on Drug Use and Health 2014 [29], there are about 72% samples about consuming cannabis in 2014, 14% consuming pills and 13% consuming cough syrup. We multiplied those percentages by the total number of posts we have for three kinds of drugs and plotted those results as orange bars in Figure 3. Therefore, as we can see from Figure 3, this survey basically is consistent with the patterns we found: weed is the most popular, the consumption of pills and cough syrup are much less.

### B. Time Patterns for Different Drugs

We used timestamp of each post to study the drug consumption patterns. Instagram records the GMT time for each post regardless of the time zone it was created in, which does introduce some errors while studying consumption patterns. Therefore, in order to verify our dataset, we searched through all the drug related activity posts that have geolocation from those 406 drug users (176 users have posts with geolocation tags) and applied the Google Map API to the location we found. Among all the 176 drug users, 172 drug users have posted drug activities inside United States. Therefore, we can assume that the majority of our posts are from United States, therefore this assumption does not introduce any significant errors from other countries' posts in our analysis. After using the timestamps to study frequency of consumption, we found a number of interesting patterns related to drug use, and these are as follows (Fig. 4 and Fig. 5).

Regarding hour of day, we see two peaks of consumption for weed, one at 16:00 hours and another at 21:00 hours. This seems legitimate because there exists a '420' culture in the weed community, which involves the act of consuming weed at 4:20 PM in the day. Also, since we have GMT timestamps, one peak corresponds to the 4:20 movement in GMT time zones, and the next peak corresponds to the 4:20 movement in the North American continent. Moreover, the consumption of drugs seems to increase towards the end of the day. Regarding day of week, people are more likely to consume drug on Thursdays than any other days (part of the reason is that people tend to have parties on Thursdays). Moreover, the consumption on Fridays is surprisingly higher than that on weekend days.

Additionally, it is likely that many Instagram users do not post their pictures or videos during the events but long time after the event, therefore our data might introduce noise due to this group of people. Let us assume our drug users post their drug activities an arbitrary amount of time after the activity. If it were true, our time distribution by the hour of day would look like Fig. 6 (the overall distribution of posts by the hour of day on Instagram) [30]. However, Fig. 4 is quite different from Fig. 6. Therefore, we can infer that most of our drug users do not post drug activities after an arbitrary amount of time from when the event happens. Therefore, for our dataset, the majority of the drug users probably post real-time posts when consuming drugs.

### C. Linguistic Analysis and Dynamic Hashtags Set

Those frequent hashtag sets are sorted based on their support. We then examine the top frequent drug-related hashtags and if the hashtag set, say, has a support over 20% (which is a relatively high support threshold in our case) and also is not in our hashtag database, then we add the hashtags in the set to our database.

During the mining process of frequent hashtag sets, we found some very interesting drug-related words or lingo. Some words that are drug-related were not used to describe drug use before, for examples:

- 'highsociety' has a support of more than 10% but was used to describe 'luxury life' instead of 'getting high'
- with support 5%, 'faded' that used to tell a feeling of 'unconsciousness' now means drug consumption
- 'poup' was meaning 'mix up' but now means 'make some cough syrup'

A common challenge with flagging sensitive posts is that language is constantly changing, with new words being added to the drug vocabulary and words having different meanings in drug related context, as shown above. Hence, dictionaries compiled quickly become outdated because of ever changing language. Using our approach, we have found a way to keep the drug related terms dictionary up to date, by adding the frequently occurring terms mined using the Apriori algorithm. However, this does involve human supervision, because some tags that appear with drug related tags are not drug related themselves, #goodtime for example. Hence, human supervision is required to select which hashtags to add to the dictionary.

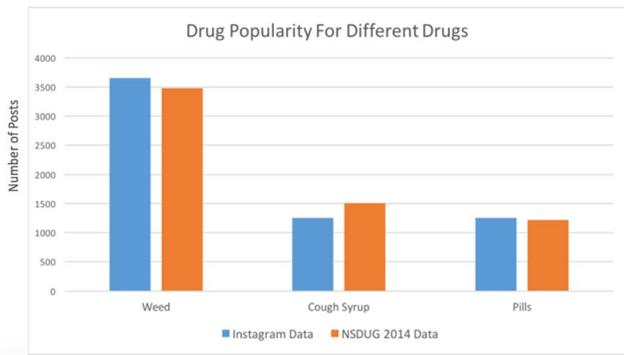

Figure 3. The comparison between the results by mining Instagram and the results from NSDUG 2014 survey data.

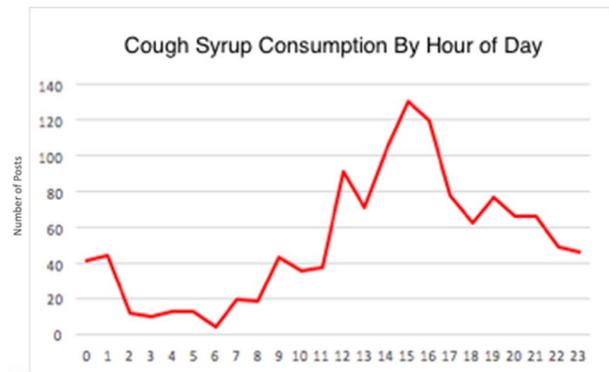

Figure 4. The time patterns of different illicit drug use (by day).

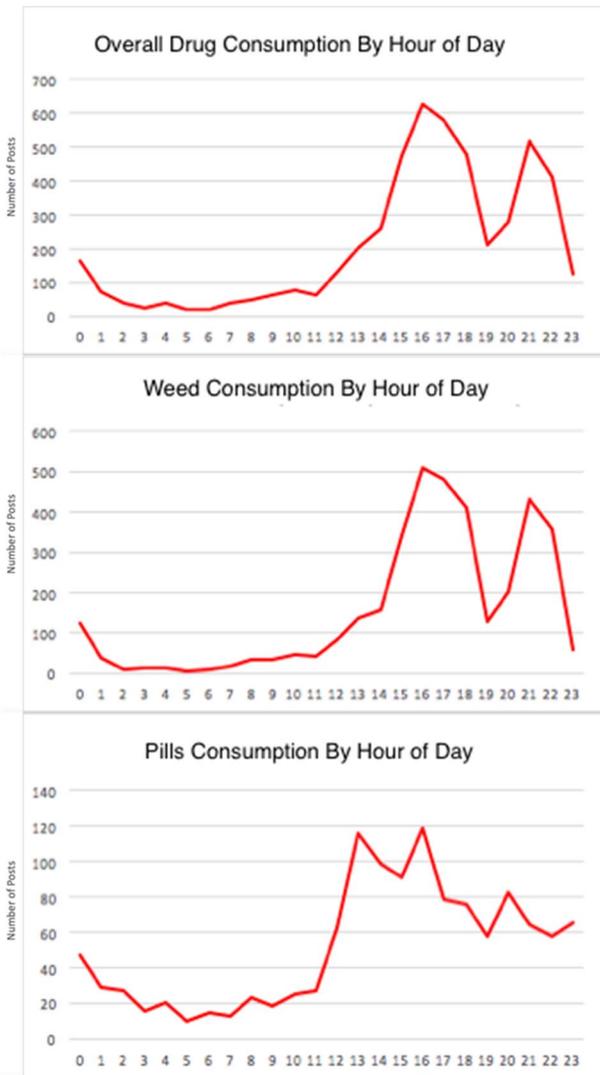

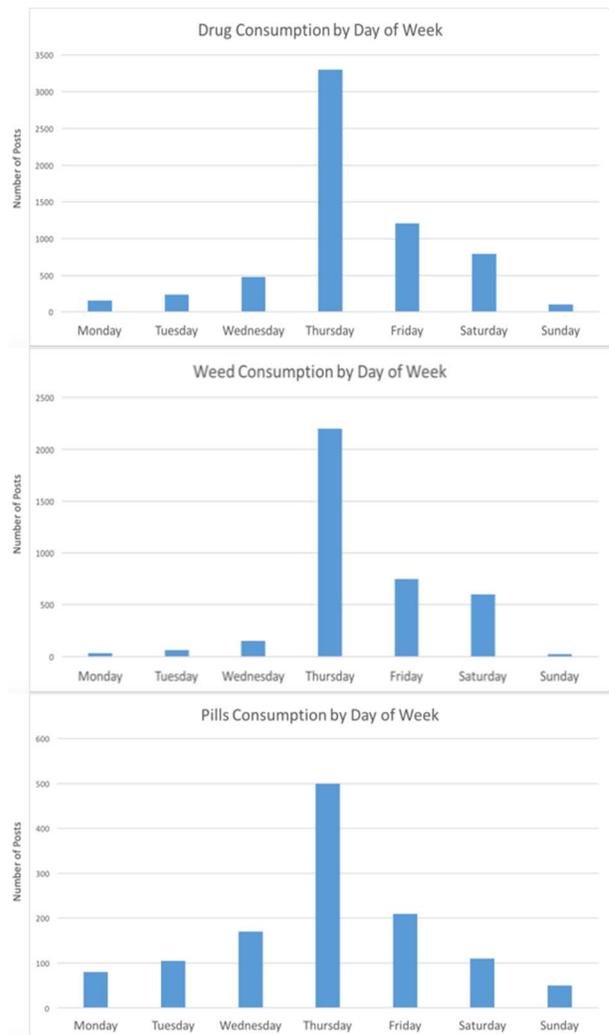

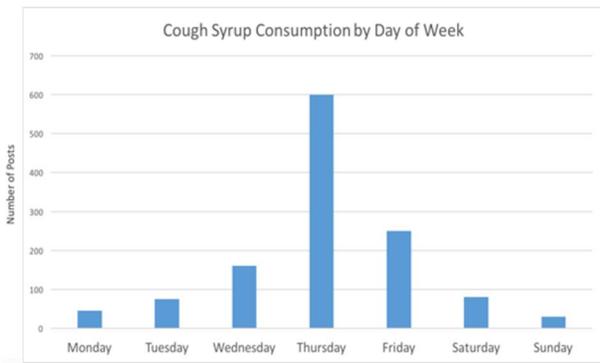

Figure 5. The time patterns of different illicit drug use (by week).

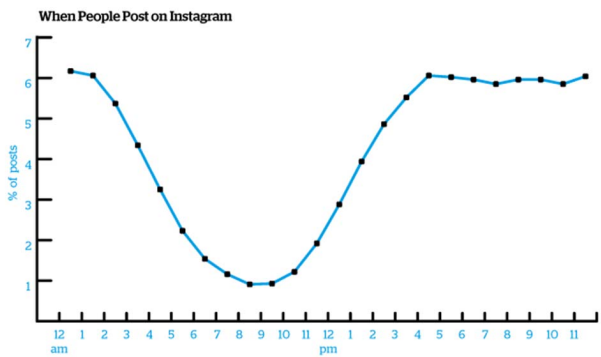

Figure 6. Overall number of Instagram posts by hour of day 2015.

### D. Image Analysis

After we obtained all the age and gender information of those 406 users, we plotted the age distribution shown in Fig. 6. Apparently, the majority of the drug users on Instagram are between 20-40. Also, there are some teenagers (15-20 years-old) who smoke cannabis, which is a potential public health concern. Moreover, comparing with the results from the National Survey on Drug Use and Health (2013), our age distribution has the similar peaks (peaks locates at young generation).

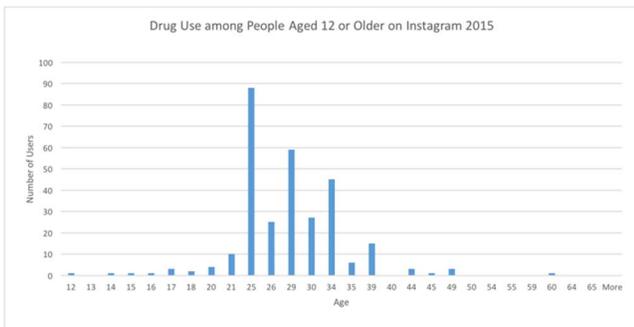

Figure 6. Drug Use among People Aged 12 or Older on Instagram 2015.

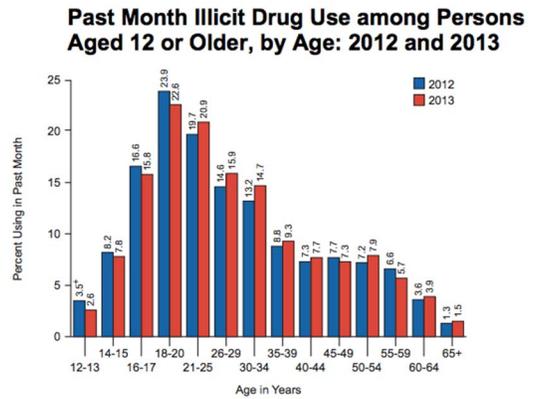

Figure 7. Past Month Illicit Drug Use among Persons Aged 12 or Older by Age: 2012 and 2013.

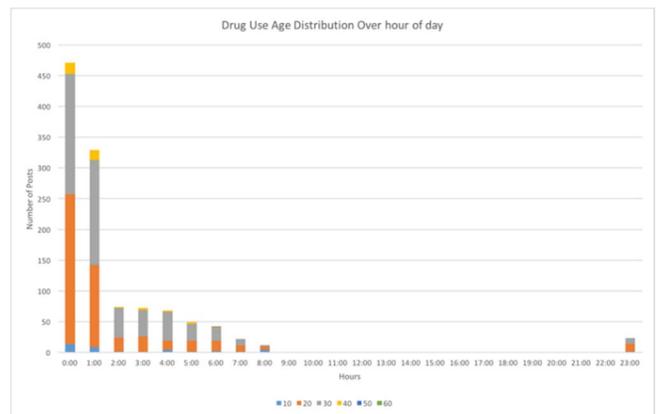

Figure 8. Drug Use Age Distribution over hour of day.

Project Oxford also provides gender detection. Among 406 drug users, 145 are detected as females and 261 are detected as males. This is also consistent with the data from the National Survey on Drug Use and Health, where illicit drug users consist of more males than females.

Finally, we plotted a stacked column chart shown in Fig. 8. There are roughly equal number of people between 20 to 30 years old (orange part) and people between 30 to 40 years old (grey part) smoking around midnight. But, from 2am to 6 am, people between 30 to 40 years old smoke more than people between 20 to 30 years old. Moreover, teenager (blue part) is more likely to smoke around midnight than any other time.

### E. Common Interests of Drug Users

Now that we have a list of drug users, we can use social media to study their likes and interests. The advantage of social media is that we can easily track the pages a user follows, and analyze them. Using the Instagram relationship endpoint, we record the pages the drug users in our paper follow, and applied A-priori (frequent item-set fetching algorithm) to the data to discover commonly followed pages and association rules.

For the top 10 frequent followed accounts, we manually check each one of them and identify whether they are common interesting pages followed by drug users. We found many pages that had very high supports and there were association rules that had a probability of 1. These are quite interesting:

- ('sdryno','coylecondenser')==>('oilbrothers','elksthatrun') with confidence 0.6. 'elksthatrun' is a music band that is followed by a lot of drug users, which may mean that drug users tend to like a certain kind of music
- ('cheechandchong')==>('heytommychong','hightimes magazine'), which implies that people who like comedian 'cheechandchong' also like 'hightimesmagazine'. Therefore, drug users are following some particular comedians.

From the above points, we see that we can look at what celebrities are popular among drug users, what kind of music and comedy they like, and so on. Moreover, by performing network analysis we can study social interactions of drug consumers versus non-drug consumers. Many of the drug users in our database are following glass-makers (people making device used to consume drug) such as 'elboglass' with support 0.129 and 'saltglass' with support 0.107

- Weed-related pages like 'weedhumor' with support 0.102 and 'hightimesmagazine' with support 0.156
- Some celebrities are followed by a decent portion of drug users such as 'christucker' (Chris Tucker), 'therock' (WWE wrestler and actor The Rock), 'heytommychong' (comedian) and 'cheechandchong' (comedian).

*F. Network Analysis*

After obtaining drug related accounts from earlier analysis, we further investigate the differences between the social networks of drug users versus those of non-drug users. Since we already have a list of 406 drug related accounts, a list of non-drug related accounts is needed. To that end, we fetched posts containing the hashtag "#instapic", which is a generic non-drug related hashtag, and tracked down the users associated with these posts. Using a simple classifier for non-drug, we obtained a list of users that are not drug-related. To be specific, what this classifier does is to search through all the hashtags in all the posts of a user and if we find one drug-related hashtag, then this user is eliminated from the non-drug users. After this cleansing process, we used the Instagram relationship endpoint to fetch the accounts the users followed and the accounts that followed the users. We then created two graph objects out of this data, one for non-drug related accounts and one for drug-related account. After that, we computed various statistical measures on these two graphs and compared the differences between them. First, we examined the vertices with the maximum in-degree in each of the networks, and the results are as follows:

1) In the drug-related accounts, the node with the maximum in-degree is kash_stack, as shown in Fig. 9

The top 10 highest in-degree accounts for both groups are:
1) Drug users: kash_stack, blownsmokeshop(dealer), bump0(user), budz.bunnny(user), solid4real_(user), stonepicks, tiendablonsh_(public page), kingraymusic, g0dfre(dealer), _nickguillen
2) Non-drug Users: wweclipsdaily, fashionworldstyle, dieta_sana_vegetariana, openweb2.0, kasssienka, mohamadbbaban, maisontitibotiashairandmakeup, deanna.kate, bushra_asg

Most of the non-drug users' frequently liked pages are about fashion, makeup and legal sellers. However, about 80% of the drug users' frequently liked pages are about drug. Therefore, there is a clear difference between the preferences of those two communities.

Moreover, we analyzed the triangles in the graph, for example, the number of three-node cycles in the graph, and checked which vertex was involved in the most triangles in both of the graphs. In the drug-related accounts network, it is theinfamouscartel (drug user and dealer). In the non drug-related accounts network, it is Domenicabruno (a popular user on Instagram). These two accounts are the pivot or hub of the two networks. The vertices they represent connect users in each network together and not surprisingly, the pivot of the drug user community is not only a drug user but also a dealer. We measured the number of 'triangles' in both communities. For the drug user network, we found 2256 cycles. For the non-drug user network, we only found 186 cycles. The difference is remarkable: drug users form a more connected and centralized community than non-drug users partly because they need this network to trade and find dealers.

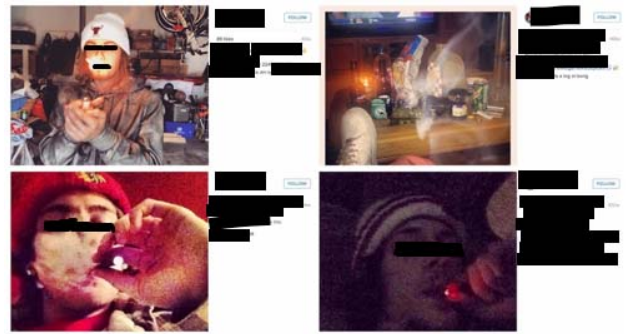

Figure 9. Screen shot of the account kash_stack

The overall centrality measures of the two networks are computed and presented in Table III. The definitions of these measures are as follows:

*In-degree: the number of arcs leading to this vertex*
*Out-degree: the number of arcs coming from this vertex*

There are some interesting observations based on Table 3. The two communities are similar in terms of followers (in-degree), but differ significantly in terms of the accounts being followed (out-degree). Drug users distinctively follow more people than non-drug users. Our hypothesis is that in addition to some commonly followed pages, drug users have a higher possibility to follow the drug dealers, glass-makers and other users to keep their supplies. To be specific, drug users need to follow more drug-related accounts to make their network and supply stable and steady.

Moreover, drug dealers have a relatively low in-degree and out-degree compared with drug users and non-drug users. We manually checked some of the dealers' accounts and found that because drug selling is a sensitive business, dealers often delete and recreate their accounts or have a number of accounts for drug selling (accounts with similar usernames, for example: chrispaul, paulochris and _paulchris).

In terms of the ratio of in/out-degree, drug dealers' accounts have a higher ratio than regular drug accounts, which means dealers are more important vertices that connect drug users' community in the network than regular drug users. Also, as we can see, even though the average degrees of drug users group and non-drug users group is similar, their in/out-degree ratios differ a lot, which means the structures of network for both groups are different. But for regular non-drug user accounts (not popular accounts, with fewer than 1000 followers) and regular drug user accounts (not popular accounts, with fewer than 1000 followers), all four columns are similar. Our guess is that popular accounts (with higher in-degree than regular users) have significant effects on the structure of the network.

## G. Drug Use Patterns by Location

After fetching all the drug related posts with geo tag from Los Angeles and Portland, we plotted those drug activities on the map, as indicated by Fig. 10.

Table III. Statistics of the drug user network and the non-drug user network.

| Group | AVE INDEGREE | AVE OUTDEGREE | AVE DEGREE | IN/OUT DEGREE RATIO |
|---|---|---|---|---|
| Drug User Group | 539.88 | 799.48 | 1073.69 | 0.68 |
| Non-Drug User Group | 554.17 | 476.10 | 1030.27 | 1.16 |
| Regular non-drug accounts(Not popular accounts) | 467.33 | 483.07 | 950.40 | 0.97 |
| Regular Drug accounts(Not popular accounts) | 495.15 | 502.94 | 998.10 | 0.98 |
| Drug Dealer | 529.36 | 463.31 | 992.68 | 1.14 |

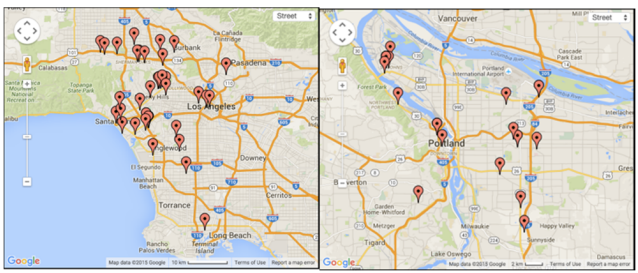

Figure 10. Pattern of drug use based on location (Los Angeles, CA and Portland, OR).

Next, we found all the clusters on the map under human supervision, which are the highly popular places to consume drug. By inputting the coordinates of these clusters into Google Maps API, we could identify those places. As a result, more than 60% of those places are residential buildings, about 10% are clubs (for example, Brentwood country club) and 30% are restaurants (including bars, for example, House of Blues, WineBar).

## VI. CONCLUSIONS AND FUTURE WORK

Instead of using the traditional methods such as surveys, we propose to leverage social media to fetch posts from drug users with significantly less time and labor cost, while achieving decent scalability, timeliness, and accuracy. First, by using hashtags-based search to find time patterns of drug consumption, we have uncovered interesting trends in the consumption of various classes of drugs. Then, by location endpoint search, we have found interesting geographical patterns and visualized the locations by drawing bubbles on the map. Moreover, in order to keep updating our database of drug-related hashtags, we applied Frequent Itemset Mining to the hashtags in drug users' posts we found by using recent media searching. Finally, relationship endpoint gives us a way to find potential network among drug users and drug-related pages on Instagram.

An important innovation of this study is in multimedia data analysis, and especially face image analysis. We employ the state-of-the-art Face API of Project Oxford by Microsoft to study the age and gender patterns of drug users. Such fine-grained demographics at a large scale are quite consistent with the findings of NSDUH [29], demonstrating the potential of image-based data analytics for studying drug use and other risky behaviors such as underage drinking [4]. In addition to age and gender, race related patterns can be discovered in a similar fashion, again facilitated by face image analysis.

Meanwhile, we have produced potentially significant results that can be utilized in areas including linguistics and psychology. For example, based on our database of drug-related hashtags, we can discover drug slang evolving patterns, for example, the hashtag '#dirtysprite' has been evolving to '#dirtysprite2', '#dirtysprites' and 'dirtyspritefoever'.

For deeper network analysis, we plan on collecting network data within two hops away from our user vertices, for example, the follows of the followers of our users, and then perform community clustering analysis on the networks to understand how communities differ between drug-related networks and non-drug related networks. We hope to use this information to create a more robust classifier to identify drug dealers and consequently design strategies to disrupt such drug related networks.

In the future, by studying the characteristics of drug use locations in Los Angeles, we hope to use a classifier to predict drug hot spots in other cities such as San Francisco. Prior studies have found that venues can be used to predict likely activities [18]. Furthermore, since drug use will cause serious influence on the mind [6], we plan to find the correlation between drug activities and crime rates in a given area.

We can further improve our analysis tools once we build classifiers that we can use to classify drug dealers, users, and glass makers/legitimate sellers, and then use network analysis to study the social interactions of drug users compared to non-drug users.

VII. ACKNOWLEDGMENTS

We gratefully acknowledge the support from the University of Rochester, New York State through the Goergen Institute for Data Science, and the valuable discussions with the New York State Attorney General's Office.

# Appendix A – Dictionary of Terms related to Illicit Drug Consumption

| | | |
|---|---|---|
| Norcos | klonopin | ganja |
| kush | clonazepam | hightime |
| weed | kpins | maryjane |
| 420 | Vicodin | shatter |
| marijuana | vicoden | topshelflife |
| wax | gg249 | highlife |
| mgp | M367 | stoned |
| dank | alprazolam2mg | bitxhimhigh |
| highsociety | benzodiazepines | actavis |
| dabs | wockhardt | wfayo |
| weedporn | purpledrink | weshouldsmoke |
| thc | promethazinecodeinest8 | mmj |
| str8drop | drop | Blunts |
| caliweed | codeinekid | Chronic |
| bxtchimhigh | purplebecomingin | Dope |
| dopehead | dirtyspritegang | Herb |
| bud | nosealnodeal | Joint |
| Meds | hydrocodonesyrup | potheadsociety |
| promethazinekings | bxtchimhigh | |
| weedstagram | tussionex | |
| ounces | codiene | |
| staylifted | promethazinekings | |
| cannabiscommunity | leanforsale | |
| hydro | sizzurp | |
| rawmuch | nodsquadd | |
| pillgame | dirtysprite2 | |
| meds | vampblood | |
| pillgang | oxycodone | |
| lortab | hydrocodone | |
| xanaxbars | norcos | |
| nodsquadd | alprazolam | |
| Painkillers | opiates | |
| mdma | 2mg | |
| smoke | benzos | |
| glock_team | percocet | |
| blazed | 30mg | |
| stone | percs | |
| dab | xans | |
| oil | xannies | |
| hash | xannys | |
| bong | vicodin | |
| vape | a215 | |
| Maryjane | r039 | |
| Munchies | Percocet | |
| bluedream | xanies | |
| juicyfruit | s903 | |
| sourdiesel | hydromorphone | |
| dankassbombass | alprolozam | |
| toke | pillpopper | |
| cannabis | Norcos | |
| pot | Oxycodone | |
| codeine | roxycodone | |
| promethazine | Zoloft | |
| doublecup | vicodins | |
| qualitest | percocets | |
| caraco | gg249 | |
| hitechred | Meds | |